\documentclass[12pt]{article}
\def\be{\begin{equation}}
\def\ee{\end{equation}}
\def\bea{\begin{eqnarray}}
\def\eea{\end{eqnarray}}
\usepackage{graphicx}

\catcode`\@=11
\def\lsim{\mathrel{\mathpalette\@versim<}}
\def\gsim{\mathrel{\mathpalette\@versim>}}
\def\@versim#1#2{\vcenter{\offinterlineskip
\ialign{$\m@th#1\hfil##\hfil$\crcr#2\crcr\sim\crcr } }}
\catcode`\@=12
\usepackage{axodraw}

\catcode`@=12
\begin{document}
\thispagestyle{empty}
\begin{flushright}
UCRHEP-T469\\
May 2009\
\end{flushright}
\vspace{0.3in}
\begin{center}
{\LARGE \bf Inverse Seesaw Neutrino Mass from\\
Lepton Triplets in the $U(1)_\Sigma$ Model\\}
\vspace{1.5in}
{\bf Ernest Ma\\}
\vspace{0.2in}
{\sl Department of Physics and Astronomy, University of California,\\ 
Riverside, California 92521, USA\\}
\end{center}
\vspace{1.5in}
\begin{abstract}\
The inverse seesaw mechanism of neutrino mass, i.e. $m_\nu \simeq (m_D^2 
/m_N^2) \epsilon_L$ where $\epsilon_L$ is small, is discussed in the context 
of the $U(1)_\Sigma$ model.  This is a gauge extension of the Standard Model 
of particle interactions with lepton triplets $(\Sigma^+,\Sigma^0,\Sigma^-)$ 
as (Type III) seesaw anchors for obtaining small Majorana neutrino masses.
\end{abstract}

\newpage
\baselineskip 24pt

\noindent \underline{\it Introduction} : If the $SU(3)_C \times SU(2)_L \times 
U(1)_Y$ Standard Model (SM) of particle interactions is extended 
\cite{flhj89,m98} to include a lepton triplet $(\Sigma^+,\Sigma^0,\Sigma^-)$ 
per family, then the heavy Majorana mass of $\Sigma^0$ acts as a seesaw anchor 
for the neutrino to acquire a small mass, just as in the case of using a 
singlet $N$.  This is often referred to as Type III seesaw \cite{type3}.  
It has also been shown \cite{m02,mr02,bd05,aem09} that this addition admits 
a new $U(1)_\Sigma$ gauge symmetry, with the following charges:
\begin{eqnarray}
(u,d)_L \sim (3,2,{1 \over 6})&:& n_1, \\u_R \sim (3,1,{2 \over 3})&:& 
n_2 = {1 \over 4}(7n_1-3n_4), \\ d_R \sim (3,1,-{1 \over 3})&:& n_3 = 
{1 \over 4}(n_1 + 3n_4), \\ (\nu,e)_L \sim (1,2,-{1 \over 2})&:& n_4, \\ 
e_R \sim (1,1,-1)&:& n_5 = {1 \over 4}(-9n_1 + 5n_4), \\ (\Sigma^+,
\Sigma^0,\Sigma^-)_R \sim (1,3,0)&:& n_6 = {1 \over 4}(3n_1 + n_4).
\end{eqnarray}

Consider now the $inverse$ seesaw mechanism: a situation is established 
where $m_\nu = 0$ because of a symmetry, which is then broken by a $small$ 
mass parameter \cite{ww83,mv86,m87,m09-1,m09-2}. As a result, $m_\nu$ is 
proportional to this symmetry-breaking parameter (and not inversely 
proportional to $m_N$ as in the usual seesaw).  The prototype model is to 
add a singlet Dirac fermion $N$, i.e. both $N_R$ and $N_L$, with lepton 
number $L=1$ per family to the SM.  The $3 \times 3$ mass matrix spanning 
($\bar{\nu}_L, N_R, \bar{N}_L$) is then given by
\begin{equation}
{\cal M}_{\nu,N} = \pmatrix{0 & m_D & 0 \cr m_D & \epsilon_R & m_N \cr 
0 & m_N & \epsilon_L},
\end{equation}
where $\epsilon_{L,R}$ are lepton-number violating Majorana mass terms. This 
is a natural extension of the famous $2 \times 2$ seesaw mass matrix, but it 
also has a clear symmetry interpretation, i.e. $\epsilon_{L,R}$ may be 
naturally small because their absence would correspond to the exact 
conservation of lepton number. [A linear combination of $\nu_L$ and $N_L$ 
would combine with $N_R$ to form a Dirac fermion, whereas its orthogonal 
combination would remain massless.]  Using $\epsilon_{L,R} << m_D, m_N$, the 
smallest mass eigenvalue of Eq.~(7) is then $(m_D^2/m_N^2) \epsilon_L$ which 
can be small because $\epsilon_L$ is small even if $m_N$ is only 1 TeV.

\noindent \underline{\it Inverse seesaw using $U(1)_\Sigma$}~:~ To implement 
the inverse seesaw in the $U(1)_\Sigma$ model, the lepton triplet
\begin{equation}
(\Sigma^+,\Sigma^0,\Sigma^-)_L \sim (1,3,0)~:~ n_7
\end{equation}
is added.  This will not change $n_{1,2,3,4,5}$, but now
\begin{equation}
n_6 - n_7 = {1 \over 4}(3n_1 + n_4) \equiv n_0
\end{equation}
is required to cancel the $[SU(2)]^2 U(1)_\Sigma$ triangle anomaly.  Using 
this, the mixed gravitational-gauge anomaly of $U(1)_\Sigma$ is also zero. 
However, the $[U(1)_\Sigma]^3$ triangle anomaly is no longer zero, unless 
either $n_7=0$ or $n_6=0$.  Extra singlets are needed to cancel this 
anomaly without affecting the others.  Details will be presented in a 
later section.

To be specific, consider the case $n_1=n_4=1$, and $n_6 = -1$, then 
$n_0 = n_2 = n_3 = 1$, $n_5 = -1$, and $n_7 = -2$.  Three Higgs doublets 
are needed:
\begin{eqnarray}
(\phi^+,\phi^0)_1 &:& n_1-n_3=n_2-n_1={3 \over 4}(n_1-n_4)=0, \\
(\phi^+,\phi^0)_2 &:& n_4-n_5={1 \over 4}(9n_1-n_4)=2, \\
(\phi^+,\phi^0)_3 &:& n_6-n_4=-{1 \over 4}(3n_1+5n_4)=-2,
\end{eqnarray}
coupling to quarks, charged leptons, and neutrinos respectively.  Note that 
$(\Sigma^+,\Sigma^0,\Sigma^-)_L$ does not couple to $(\nu,e)_L$ because 
$-n_7 -n_4 = 2n_0 - n_4 = 1$ and there is no Higgs doublet with that charge.
Add scalar singlets with $U(1)_\Sigma$ charges as follows:
\begin{equation}
\chi_1 \sim -1, ~~~ \chi_2 \sim 2, ~~~ \chi_4 \sim -4,
\end{equation}
then the term $\bar{\Sigma}_L \Sigma_R \chi_1$ provides a Dirac mass $m_\Sigma$ 
linking $\Sigma_L$ with $\Sigma_R$ and the analog of Eq.~(7) is
\begin{equation}
{\cal M}_{\nu,\Sigma} = \pmatrix{0 & m_D & 0 \cr m_D & m_2 & m_\Sigma \cr 
0 & m_\Sigma & m_4},
\end{equation}
where $m_D$ comes from $\langle \phi^0_3 \rangle$, $m_\Sigma$ from 
$\langle \chi_1 \rangle$, $m_2$ from $\langle \chi_2 \rangle$, and 
$m_4$ from $\langle \chi_4 \rangle$.

\noindent \underline{Higgs structure of $U(1)_\Sigma$}~:~ To understand why 
$m_4$ can be very small in Eq.~(14), consider the Higgs potential involving 
$\chi_{1,2,4}$:
\begin{eqnarray}
V_\chi = \sum_i \mu^2_i \chi_i^\dagger \chi_i + {1 \over 2} \sum_{i,j} 
\lambda_{ij} (\chi_i^\dagger \chi_i)(\chi_j^\dagger \chi_j) + 
[\mu_{21} \chi_2 \chi_1^2 + \mu_{42} \chi_4 \chi_2^2 + \lambda_{124} \chi_1^2 
\chi_2^\dagger \chi_4^\dagger + H.c.],
\end{eqnarray}
where $\lambda_{ij} = \lambda_{ji}$.  Let $\langle \chi_i \rangle = u_i$, then 
the conditions for $V_\chi$ to be at 
its minimum are
\begin{eqnarray}
&& u_1 [\mu_1^2 + \lambda_{11} u_1^2 + \lambda_{12} u_2^2 + \lambda_{14} u_4^2 
+ 2 \mu_{21} u_2 + 2 \lambda_{124} u_2 u_4] = 0, \\
&& u_2 [\mu_2^2 + \lambda_{12} u_1^2 + \lambda_{22} u_2^2 + \lambda_{24} u_4^2 
+ 2 \mu_{42} u_4] + \mu_{21} u_1^2  + \lambda_{124} u_1^2 u_4 = 0, \\
&& u_4 [\mu_4^2 + \lambda_{14} u_1^2 + \lambda_{24} u_2^2 + \lambda_{44} u_4^2] 
+ \mu_{42} u_2^2 + \lambda_{124} u_1^2 u_2 = 0.
\end{eqnarray}
A natural solution exists \cite{ms98,m01,glr09}, such that $u_4 << u_2 << u_1$, 
i.e.
\begin{equation}
u^2_1 \simeq {-\mu_1^2 \over \lambda_{11}}, ~~~ u_2 \simeq {-\mu_{21} u_1^2 
\over \mu_2^2 + \lambda_{12} u_1^2}, ~~~ u_4 \simeq {- u_2 (\mu_{42} u_2 + 
\lambda_{124} u_1^2) \over \mu_4^2 + \lambda_{14} u_1^2}.
\end{equation}
If $\mu_{21}=0$ and $\lambda_{124}=0$, then an extra global U(1) symmetry 
exists for $\chi_1$ in Eq.~(15). If $\mu_{42}=0$ and $\lambda_{124}=0$, then 
the same holds for $\chi_4$.  Hence it may be argued that $\mu_{21}$, 
$\mu_{42}$, and $\lambda_{124}$ are naturally small. For example, let $u_1 
\sim \mu_2 \sim \mu_4 \sim m_\Sigma \sim 1$ TeV, $\mu_{21} \sim \mu_{42} 
\sim 1$ GeV, and $\lambda_{124} \sim 10^{-6}$, then $u_2 \sim 1$ GeV, $u_4 
\sim 1$ keV, and  $m_\nu \simeq (m_D^2/m^2_\Sigma)m_4 \sim 0.1$ eV for $m_4 \sim 
u_4 \sim 1$ keV and $m_D \sim 10$ GeV. The mixing between $\nu_L$ and 
$\Sigma^0_L$ is given by $m_D/m_\Sigma \sim 10^{-2}$ (and that between $e^-_L$ 
and $\Sigma^-_L$ by $\sqrt{2} m_D/m_\Sigma$) which is not constrained 
directly by the neutrino mass and may well be observable.

With the addition of $\Phi_{1,2,3}$, there are the following allowed terms: 
\begin{equation}
\Phi_1^\dagger \Phi_2 \Phi_1^\dagger \Phi_3, ~~ \Phi_1^\dagger \Phi_2 
\chi_2^\dagger, ~~ \Phi_1^\dagger \Phi_2 \chi_1^2, ~~ \Phi_1^\dagger \Phi_3 
\chi_2, ~~ \Phi_1^\dagger \Phi_3 (\chi_1^\dagger)^2, ~~ \Phi_2^\dagger \Phi_3 
\chi_4^\dagger, ~~ \Phi_2^\dagger \Phi_3 \chi_2^2.
\end{equation}
This shows that there is no extra unwanted U(1) global symmetry. Since 
$\Phi_2$ and $\Phi_3$ have opposite charges under $U(1)_\Sigma$,  
$Z-Z'_\Sigma$ mixing is zero if $v_2=v_3$ is imposed.

\noindent \underline{\it Additional fermion singlets and dark matter}~:~ 
The $[U(1)_\Sigma]^3$ triangle anomaly per family is given by
\begin{equation}
3n_0^3 - 3n_6^3 + 3n_7^3 = -9n_0 n_6 n_7 = -18,
\end{equation}
where Eq.~(9) has been used.  To cancel it, add $(2,4,2)$ copies of 
left-handed fermion singlets $(S_1,S_3,S_5)$ of charges $(1/2,-3/2,5/2)$ 
respectively. Then $2(1/2)^3 + 4(-3/2)^3 + 2(5/2)^3 = 18$ and 
$2(1/2) + 4(-3/2) + 2(5/2) = 0$, as desired.
The $S_i$ fermions do not couple to $(\nu,e)_L$ through any of the three Higgs 
doublets because of their $U(1)_\Sigma$ charges.  This is important, otherwise 
$m_\nu$ will have contributions from them as well.  They do couple among 
themselves through $\chi_1$, i.e. $S_1 S_1 \chi_1$, $S_1 S_3 \chi^\dagger$, 
and $S_3 S_5 \chi_1$.  This means that all $S_i$ fermions are heavy at the 
$U(1)_\Sigma$ breaking scale.  The lightest among them is then a possible 
dark-matter candidate, having a conserved residual parity 
\cite{hln08,l08,m08-1,m08-2} from $U(1)_\Sigma$ breaking.

\noindent \underline{\it $U(1)_\Sigma$ phenomenology}~:~ There are two salient 
features of the proposed $U(1)_\Sigma$ model.  One is the new $Z'_\Sigma$ 
boson, which is observable at the Large Hadron Collider (LHC) if kinematically 
allowed \cite{mr02,aem09}.  The other is the Dirac lepton triplet $(\Sigma^+,
\Sigma^0,\Sigma^-)$, as opposed to the usual Majorana triplet of the Type III 
seesaw \cite{type3}.  Instead of just one kind of heavy charged lepton 
$\Sigma^\pm$ with mass $m_\Sigma$ \cite{m05,bs07,df07,bns07,f07,fhs08,abbgh08,
ff08,moy08,ms09,aa09-1,m08-3,ho09,abc09,abghhps09}, there are two kinds 
\cite{aa09-2,a09} in this model with (essentially) the same mass.  Let 
$\Sigma^+_{L,R}$ be called $\Sigma_1^+$ and $\Sigma^-_{L,R}$ be called 
$\Sigma^-_2$, then $\Sigma_1^\pm$ has $L=\pm 1$, whereas $\Sigma_2^\pm$ has 
$L=\mp 1$.  Hence only $\Sigma_2^-$ mixes significantly with the usual 
charged lepton $e^-$ and $\Sigma^0$ with $\nu$.  Consequently, $\Sigma_2^-$ 
will decay into both $e^- Z$ and $\nu W^-$, but $\Sigma_1^+$ only into 
$\nu W^+$.  Unfortunately, since they have the same mass, they are not 
distinguishable at the LHC or a linear $e^+ e^-$ collider.  On the other hand, 
in an $e^-p$ collider, only $\Sigma_2^-$ and $\Sigma_1^+$ 
are produced. They may thus be distinguished by their decay modes. 
 
\noindent \underline{\it Conclusion}~:~ The inverse seesaw mechanism of 
neutrino mass is very much suited for measurable effects at the TeV scale, 
through the mixing of new heavy fermions with the known leptons.  In the 
usual seesaw, this mixing is constrained by the smallness of the neutrino 
mass and may well be too negligible for it to be observed.  In the proposed 
$U(1)_\Sigma$ model, with a new $Z'_\Sigma$ boson as well as three Dirac 
lepton triplets $(\Sigma^+,\Sigma^0,\Sigma^-)$, a rich phenomenology is 
expected at the LHC and beyond.

\noindent \underline{\it Acknowledgement}~:~ 
This work was supported in part by the U.~S.~Department of Energy under
Grant No. DE-FG03-94ER40837.

\baselineskip 16pt
\bibliographystyle{unsrt}

\end{document}